\documentclass[onecolumn,manuscript]{revtex4}
\usepackage{amssymb}

\usepackage{epsfig}
\usepackage[english]{babel}
\usepackage{latexsym}
\usepackage{graphics}
\usepackage{subfigure}
\usepackage{epsfig}
\usepackage{graphicx}
\usepackage{dcolumn}
\usepackage{amsmath}

\def\hJ{{\displaystyle{\cal J }}}

\begin{document}                

\title{Imaging molecular characterization by high-order above threshold ionization in intense laser field}

\author{Bingbing Wang$^{1*}$, Yingchun Guo$^{2}$, Zong-Chao Yan$^{3, 4}$ and Panming Fu$^{1}$}

\address{$^1$Laboratory of Optical Physics, Beijing National Laboratory for Condensed Matter Physics, Institute of
Physics, Chinese Academy of Sciences, Beijing 100080, China}
\address{$^2$State Key Laboratory of Precision Spectroscopy,
East China Normal University, Shanghai 200062, China}
\address{$^3$Department of Physics, University of New Brunswick, P. O. Box
4400, Fredericton, New Brunswick, Canada E3B 5A3}
\address{$^4$Center for Theoretical Atomic and Molecular Physics, the Academy of Fundamental
and Interdisciplinary Sciences,
Harbin Institute of Technology, Harbin 150080, China}

\begin{abstract}
Using the frequency-domain theory, the high-order above threshold
ionization (HATI) of a molecule is treated as a two-step
transition: an above threshold ionization (ATI) followed by a
laser-assisted collision (LAC). It is found that the HATI spectrum
carries three pieces of information: the fingerprint of the
molecular wave function lying in the ATI transition, the geometry
structure of the molecule lying in the potential scattering
between two plane waves, and the interaction between the ionized
electron and the laser field lying in a phase factor in the LAC
transition, where the phase factor can be regarded as due to the
action difference of the classical trajectory before and after the
collision within one laser cycle. Based on this picture, it is
demonstrated that the skeleton structure of a molecule can be
imaged more clearly using a relatively higher frequency laser
field due to the simplification of interference originated from
the phase factor. This work can shed light on imaging the
structure of a complex molecule by a HATI process.

\end{abstract}

\par\noindent

\maketitle

\par\noindent
$^*$ Author to whom correspondence should be addressed.

Laser-driven rescattering process of electrons opens a new way to
image the structure of a molecule. Both high-order harmonic
generation (HHG)~\cite{1,2,3,4,5,6} and high-order above threshold
ionization (HATI)~\cite{7,8,9} processes have been utilized to
realize this task. Odzak and Milosevic~\cite{6} have shown clear
minima on HHG spectra that depend on molecular
orientation. Becker \emph{et al.}~\cite{8} have demonstrated that the
interference structure of a HATI spectrum strongly depends on
the symmetry of the ground state molecular orbital. A general
formula that describes the destructive interference of a
multiple-center recollsion has been found in both HHG~\cite{6} and
HATI~\cite{8} spectra. Now people are trying to extract more
information about a molecule from the HATI spectrum. Therefore,
it is necessary to understand in more detail how a complex
interference pattern forms in the process, and what characteristic
of the whole system the pattern represents.

The frequency-domain theory based on a nonperturbative quantum
electrodynamics was developed by Guo~\emph{et al.}~\cite{10} in
1989 dealing with the above threshold ionization (ATI). Then,
Gao~\emph{et al.}~\cite{11} extended this method to a study of HHG
process. Fu \emph{et al.}~\cite{12,13} further explained the
origin of the plateau structure on HHG spectrum and demonstrated
the relationship between the time-domain and frequency-domain
theories in strong laser physics. More recently, Wang~\emph{et
al.}~\cite{14} applied this theory to investigate HATI process. It
should be noticed that the characteristic of the frequency-domain
theory is that the time-evolution dynamic process of a quantum
system in a strong laser field is mapped into a frequency space;
hence all the rescattering processes can be treated as a two-step
transition. Especially, the HATI of a molecule in a strong laser
field can be treated as an ATI followed by a laser-assisted
collision (LAC). Consequently, we find in this paper that the
characteristic of a molecule-laser system can be identified from
the HATI spectrum: the fingerprint of the molecular wave function
lies in the ATI transition, the geometry structure of the molecule
lies in the potential scattering term between two plane waves, and
the interaction between the ionized electron and the laser field
lies in a phase factor in the laser-assisted transition.
Furthermore, since these three parts are relatively independent
with each other, we demonstrate that the molecular skeleton
structure can be imaged more clearly with a higher frequency laser
field because of the simplification of interference due to the
interaction between the ionized electron and the laser field.
Recently, we are aware of the work by Reiss on the ATI of an atom
in a low frequency laser field using the strong-field
approximation~\cite{reiss}.

The frequency-domain theory of the HATI process has been described in \cite{14}.
Here we summarize the basic elements of this theory and the
modifications for a molecular case. Atomic units are used
throughout unless otherwise stated. We consider a quantized
single-mode laser field of frequency $\omega$. The Hamiltonian for
a molecule-laser system is
\begin{eqnarray}
H=H_0+U({\bf r})+V, \label{e1}
\end{eqnarray}
where $ H_0={ (-i\bigtriangledown)^2 \over 2m_e} + \omega N_a$ is
the energy operator for a free electron-photon system,
$N_a=(a^{\dag}a+ aa^{\dag})/2$ is the photon number operator with
$a$ ($a^{\dag}$) being the annihilation (creation) operator of the
laser photon mode, $U({\bf r})$ is the molecular binding
potential that characterizes the geometry structure of the
molecule, and $V$ is the electron-photon interaction $V=-{e \over
m_e}{\bf A}({\bf r}) \cdot (-i\bigtriangledown) +{e^2{\bf A}^2
({\bf r})\over 2m_e},$ in which ${\bf A}({\bf r}) =g(\hat{\epsilon}\,
a\, e^{i{\bf k}\cdot {\bf r}} +{\rm c.c.})$ is the vector
potential, and $g=(2\omega V_e)^{-1/2}$ with $V_e$ being the
normalization volume of the field, and $\hat{\epsilon}$ the
polarization vector of the laser field.

The time-independent feature of the field-quantized Hamiltonian
enables us to treat HATI as a genuine scattering process in an
isolated system that consists of the photons and the molecule.
Since the energy is conserved throughout the interaction, the
formal scattering theory~\cite{15} can thus be applied. Therefore,
the transition matrix element can be written as~\cite{10}
\begin{eqnarray}
T_{fi}&& =\langle\psi_f\mid V\mid \psi_i\rangle + \langle\psi_f
\mid U {1 \over E_f-H + i\epsilon} V \mid \psi_i\rangle,
\label{e4}
\end{eqnarray}
where the initial state $|\psi_i \rangle = |\Phi_i({\bf r}),n_i
\rangle = \Phi_i({\bf r}) \otimes | n_i \rangle$ is the eigenstate
of the Hamiltonian $H_0+U({\bf r})$ with the associated energy
$E_i = -E_B+(n_i+{1 \over 2})\omega$. Moreover, $\Phi_i({\bf
r})$ is the ground-state wave function of the molecular electron
with the binding energy $E_B$ and $| n_i \rangle$ is the Fock
state of the laser mode with the photon number $n_i$. Finally,
the final state $| \psi_f \rangle = | \Psi_{{\bf
p}_fn_f}\rangle$ of energy $E_f = E_{{\bf p}_fn_f}$ is the Volkov
state of the quantized-field~\cite{10,14}.

The first and second terms in Eq.(\ref{e4}) respectively correspond to the
processes of direct and rescattering ATI. Thus, $T_{fi}$ can be
expressed as $T_{fi} = T_{\rm d} + T_{\rm r}$. On the one hand, as
for the direct ATI transition, we have~\cite{10,16}
\begin{eqnarray}
T_{\rm d}&&=\langle\psi_f\mid V\mid \psi_i\rangle\nonumber \\
&&=V_e^{-1/2}\omega (u_p-j)\Phi ({\bf p}_i) \hJ_j(\zeta_f, \eta),
\label{e6}
\end{eqnarray}
where $u_p=U_p/\omega$ with $U_p$ the ponderomotive energy of an
electron in the laser field, $j=n_i-n_f$, $\Phi ({\bf p}_i)$ is
the Fourier transform of the initial wave function $\Phi_i ({\bf
r})$, and $\hJ_j (\zeta_f,\eta)=\sum_{m=-\infty}^{\infty}
J_{-j-2m}(\zeta_f)J_m(\eta)$ is the generalized Bessel function
with $\zeta_f=2\sqrt{u_p/\omega}\,{\bf p}_f \cdot \hat{\bf
\epsilon}$, ${\bf p}_f $ the momentum of the electron in the
finial state, and $\eta= u_p/2$. In general, a molecular
ground-state wave function can be constructed using the linear
combination of atomic orbital (LCAO) approximation as
in~\cite{8,17,18}. Since Eq.~(\ref{e6}) contains a molecular wave
function, the direct ATI spectrum can thus image the
characteristic of a molecular wave function. As an example, we
calculate the angle-resolved direct ATI spectra for O$_2$ and
N$_2$ molecule with the molecule axis along the laser polarization
and the laser frequency $\omega =0.114$ and $U_p=2\omega$.  The
angle $\theta_f$ is between the final momentum of the ionized
electron and the molecular axis. We find that for N$_2$ the ATI
rate decreases with $\theta_f$ . For O$_2$, the ATI rate is zero
at $\theta_f=0$. It then increases rapidly with $\theta_f$, gets
its maximum at about 20 degrees, and then decreases with
$\theta_f$. The dependence of the ATI rate on the angle agrees
well with the results by Chen \emph{et al.}~\cite{18}.

On the other hand, the rescattering ATI transition can be written
as
\begin{eqnarray}
T_{\rm r} &&= -i \pi \sum_{{\bf p}_1 n_1} \langle\Psi_{{\bf
p}_fn_f} \mid  U \mid  \Psi_{{\bf p}_1 n_1} \rangle\nonumber\\
&&\times \langle \Psi_{{\bf p}_1
n_1} \mid V \mid \Phi_i,n_i \rangle \delta(E_f - E_{{\bf p}_1 n_1}) \nonumber\\
&&=-i \pi \sum_{{\rm all\ channels}}T_{LAC}T_{ATI} \delta(E_f -
E_{{\bf p}_1 n_1})\,. \label{e5}
\end{eqnarray}
To obtain Eq.~(\ref{e5}), we used the completeness relation of the
Volkov states $|\Psi_{{\bf p}_1 n_1} \rangle$ and assumed that the
effect of the binding potential $U$ can be neglected when the
electron is in the continuum. The physics underlying
Eq.~(\ref{e5}) is clear. Specifically, $T_{ATI}=\langle \Psi_{{\bf
p}_1 n_1} \mid V \mid \Phi_i,n_i \rangle$ represents the direct
ATI amplitude, where the ground state electron absorbs $n_i - n_1$
photons from the laser field and ionizes; whereas
$T_{LAC}=\langle\Psi_{{\bf p}_fn_f} \mid U \mid \Psi_{{\bf p}_1
n_1} \rangle $ represents the amplitude of LAC in which the
ionized electron absorbs $n_1-n_f$ photons from the field during
its collision with the nucleus, resulting in the change of the
canonical momentum of the electron from ${\bf p}_1 $ to ${\bf
p}_f$. Therefore, from the frequency-domain viewpoint, the
rescattering ATI can be described simply as an ATI followed by a
LAC with all ATI channels summed up coherently. Furthermore,
Eq.~(\ref{e5}) indicates that $T_{ATI}$ provides a weight
amplitude for $T_{LAC}$ transition in each ATI channel. Because
the initial molecular wave function only appears in the $T_{ATI}$
term, it can only influence the amplitude of LAC transition,
resulting in that the wave function can affects the HATI amplitude
of each channel, rather than the structure of the HATI spectrum.
This finding can be confirmed by the HATI spectra shown
in~\cite{8}.

\begin{figure}[t]
\includegraphics[width=\columnwidth]{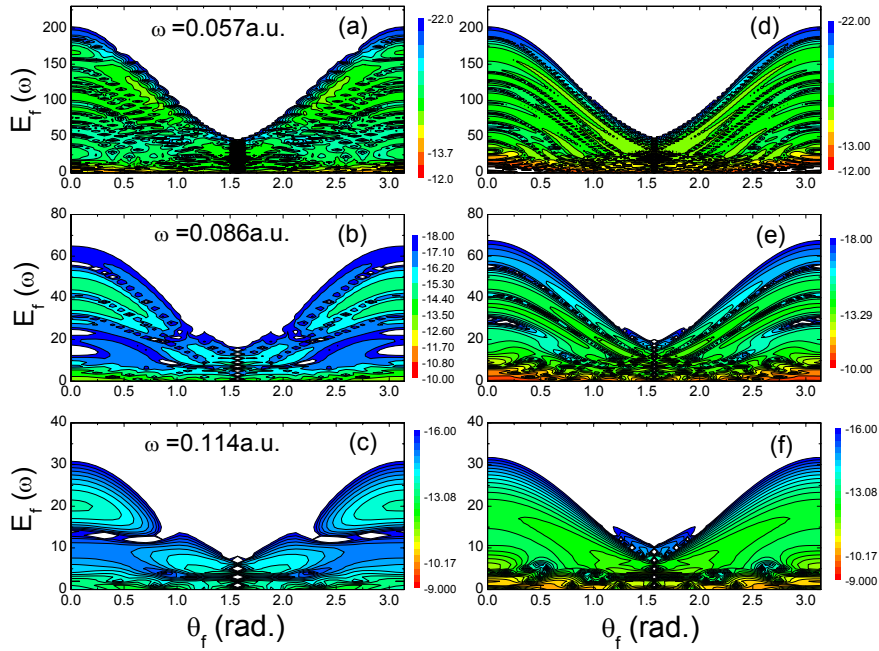} \vspace{-0.2cm}
\caption{(Color online) Angle-resolved HATI spectra with frequency
$\omega =0.057$ ((a) and (d)), $0.086$ ((b) and (e)), and $0.114$
((c) and (f)) for H$_2^+$ molecule ((a)-(c)) and Xe atom
((d)-(f)). The results are plotted in log scale.} \label{f1}
\end{figure}

The ATI transition $T_{ATI}$ in Eq.~(\ref{e5}) has the same form
as Eq.~(\ref{e6}) with ${\bf p}_f$ and $n_f$ being replaced by
${\bf p}_1$ and $n_1$, respectively. Furthermore, the transition
matrix element of LAC can be written as
\begin{eqnarray}
T_{\rm LAC} &&= \langle\Psi_{{\bf p}_fn_f} \mid  U \mid  \Psi_{{\bf p}_1n_1} \rangle  \nonumber \\
&&= V_e^{-1}J_s(\zeta_1-\zeta_f) \langle {\bf p}_f\mid U \mid {\bf
p}_1\rangle \nonumber \\
&&= V_e^{-1}\frac{\omega}{2 \pi}\langle {\bf p}_f\mid U \mid {\bf
p}_1\rangle \nonumber \\
&&\times \int_{0}^{T} dt \exp[-i(s\omega t+(\zeta_1-\zeta_f)\sin
\omega t )] ,
\label{e7}
\end{eqnarray}
where $J_s$ is the Bessel function with $s= n_1-n_f$ and $\zeta_1=
2\sqrt {u_p/\omega}\,{\bf p}_1 \cdot \hat{\bf \epsilon}$. To
obtain the last equation, we use the integral representation of
the Bessel function
$J_n(z)=\frac{1}{2\pi}\int_{-\pi}^{\pi}\exp{[-i(n\theta-z\sin\theta)]}d\theta$.
The transition term is $ \langle {\bf p}_f\mid U \mid {\bf
p}_1\rangle = \int d{\bf r} \exp[-i({\bf p}_f - {\bf p}_1)\cdot
{\bf r}] U({\bf r}) $. Equation~(\ref{e7}) indicates that the LAC
can be understood as the potential scattering between two plane
waves, as shown by $\langle {\bf p}_f\mid U \mid {\bf
p}_1\rangle$, with a phase factor $\triangle \Theta = \int_{0}^{T}
dt \exp[-i(s\omega t+(\zeta_1-\zeta_f)\sin \omega t )] $
reflecting the phase difference within one optical cycle
$T=2\pi/\omega$ between the two states before and after the
collision. The existence of the phase factor is due to the fact
that the recolliding electron oscillates by the laser field during
the recollision process. Thus, the interference pattern of the
molecular HATI spectrum can be attributed to the combination of
the potential scattering and the phase factor $\Delta \Theta$. In
particular, the skeleton structure of the molecule lies in the
potential scattering term, while the interaction between the
ionized electron and the laser field is represented by the phase
factor term in Eq.~(\ref{e7}). Since the phase factor is
independent of the potential scattering term and is only
determined by the laser condition, the phase factor just provides
a ``background" of interference fringes. Based on this
perspective, one may expect that the structure of a molecule can
be imaged more clearly by the HATI spectrum if the ``background"
interference pattern from the phase factor can be simplified.
Fortunately, this idea can be realized when the laser frequency
increases, as shown in Fig.\ref{f1}.

We now focus on the imaging of the geometry structure of a
molecule, rather than its wave function, by HATI spectrum. We
consider a HATI process of H$_2^+$ in an intense laser field. We
employ a zero-range potential model for $U(\bf{r})$ similar to
that in~\cite{19}, where the zero-range potential is $U_0({\bf r})
= \frac {2\pi} {\kappa} \delta({\bf r})\frac{\partial}{\partial
r}r,$ with $\kappa = \sqrt{2|E_B|}$. The two-center binding
potential of a diatomic molecule can thus be written as $U({\bf
r})=U_0({\bf r-z}_0) + U_0 ({\bf r+z}_0),$ where $-{\bf z}_0$ and
$+{\bf z}_0$ are the positions of the two nuclei. The molecular
orientation is along the polarization of the laser field and the
ionization potential is 29.8 eV that corresponds to the ionization
potential of the ground state of H$_2^+$ with the internuclear
separation $R_0$ being 2 a.u. The wave function of H$_2^+$ is
$\Phi_i({\bf r})=\frac{1}{\sqrt{2(1+C)}}[\phi({\bf r}-{\bf
z}_0)+\phi({\bf r}+{\bf z}_0)]$, where $\phi({\bf r})$ is the
atomic wave function corresponding to the potential $U_0({\bf r})$
and $C=\int \phi({\bf r}-{\bf z}_0)\phi({\bf r}+{\bf z}_0)d{\bf
r}$ is the atomic orbital overlap integral. The laser intensity is
$4.7\times 10^{14}$~W/cm$^2$ and the frequency is chosen to be
three different values. Figure~\ref{f1} presents the
angle-resolved HATI spectra with $\omega =0.057$ ((a) and (d)),
$0.086$ ((b) and (e)), and $0.114$ ((c) and (f)) for the H$_2^+$
molecule ((a)-(c)) and the atomic xenon ((d)-(f)). The
ponderomotive energy $U_p$ is 16.0$\omega$, 4.7$\omega$, and
2.0$\omega$ for the case of $\omega =0.057$, $0.086$, and $0.114$,
respectively. In Fig.\ref{f1}, $\theta_f$ is the angle between the
final momentum of the ionized electron and the molecular axis. It
shows that the cutoff of the kinetic energy spectrum at
$\theta_f=0$ is about 10$U_p$ for all the three frequency cases.
Comparing the molecular case with the atomic case, one can find
that (1) there are common interference fringes, which we define as
``background" interference due to $\Delta\Theta$ in both the
molecular and atomic cases, and these fringes reduce as the laser
frequency increases. (2) There exist two destructive curves for
the molecular case in Fig.~\ref{f1} (a-c), which predict the
minimum positions caused by the destructive interference of the
two-center collision, as mentioned in~\cite{8,9}. (3) The two
destructive curves in the molecular case do not change with
frequency, leading to a much clearer destructive pattern on the
HATI spectrum for the case of higher frequency, as shown in
Fig.~\ref{f1} (c).

\begin{figure}[t]
\includegraphics[width= \columnwidth]{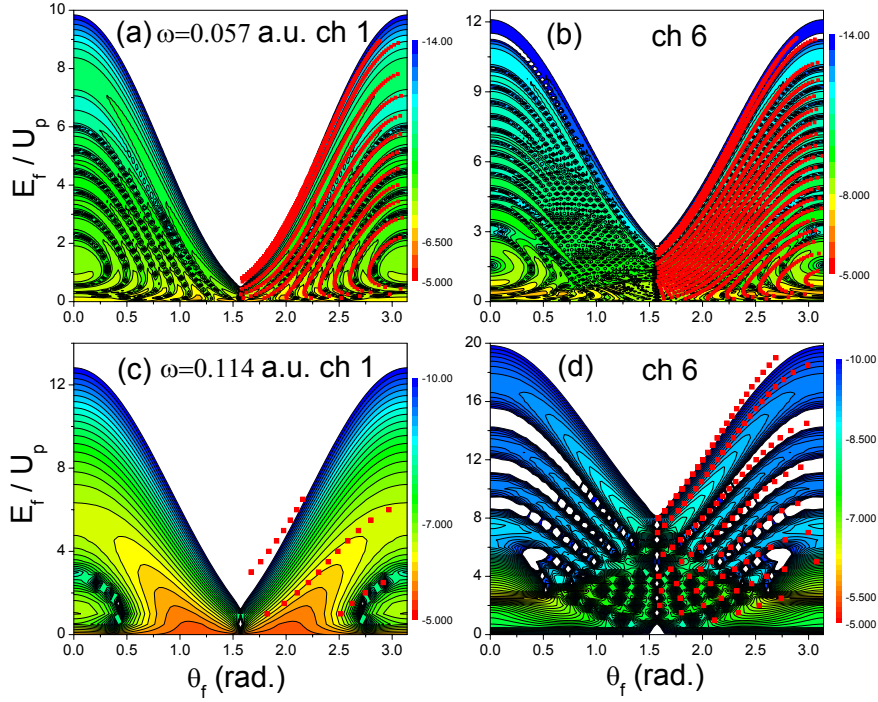} \vspace{-0.2cm}
\caption{(Color online) Angle-resolved HATI spectra of Xe with
$\omega =0.057$ ((a) and (b)) and $0.114$ ((c) and (d)) for
channel 1 ((a) and (c)) and channel 6 ((b) and (d)). The dotted
curve predicts the destructive interference fringes by setting the
cosine function Eq.~(\ref{e9}) to be zero. The results are plotted
in log scale.}\label{f2}
\end{figure}

We first investigate the source of the ``background" interference
stripes and the reason why it decreases with frequency. Let us
analyze the LAC transition for separated ATI channels. By using
the saddle point approximation, Eq.~(\ref{e7}) can be expressed as
follows:
\begin{eqnarray}
T_{\rm LAC}=&&{\omega^2 \over \pi V_e} \langle {\bf p}_f\mid U
\mid {\bf p}_1\rangle \sqrt {2\pi \over \omega^2
(\zeta_1-\zeta_f)\sin
\omega t_1} \nonumber\\
&&\times \cos [s\omega t_1 + (\zeta_1-\zeta_f)\sin \omega t_1 -
\pi/4],
\label{e8}
\end{eqnarray}
where $t_1$ is the saddle point time satisfying $\cos(\omega
t_1)=(p_f^2-p_1^2)/[4\sqrt{u_p\omega}\,({\bf p}_1-{\bf p}_f) \cdot
\hat{\bf \epsilon}]$, and $s = (p_f^2 -p_1^2)/ (2\omega)$ is the
number of photons absorbed during the collision. We can find that
the interference comes mainly from the cosine function
\begin{eqnarray}
\cos F\equiv \cos [s\omega t_1 + (\zeta_1-\zeta_f)\sin \omega t_1 -
\pi/4].
\label{e9}
\end{eqnarray}
The destructive interference lines should occur when
$F=(2n+1)\pi/2$. Figure~\ref{f2} presents the angle-resolved HATI
spectra of the Xe atom when $\omega=0.057$ ((a) and (b)) and
$\omega=0.114$ ((c) and (d)) for channel 1 ((a) and (c))  and
channel 6 ((b) and (d)) respectively. One can find that the
interference fringes increase with channel order and decrease with
frequency. The dashed lines in each panel of Fig.~\ref{f2} is
obtained by the formula $F=(2n+1)\pi/2$. For comparison, we only
present the curves with $\theta_f\in ({\pi/2, \pi}$) in
Fig.~\ref{f2}. It indicates that the dashed curves by
Eq.~(\ref{e9}) agree well with the quantum calculations for all
the cases shown in Fig.~\ref{f2}, except that the curves for the
small $E_f$ part in Fig.~\ref{f2} (b) are a little lower than the
quantum results. This difference may be due to the error from the
saddle-point approximation.

As mentioned in~\cite{14}, the phase factor in Eq.~(\ref{e7}) can
be regarded as due to the classical action difference of the
classical trajectory before and after the collision, \emph{i.e.},
$\triangle \Theta=\int_{0}^{T} dt \exp[-i\triangle S(t,{\bf
p}_1,{\bf p}_f)]$ with $\triangle S(t,{\bf p}_1, {\bf p}_f) =
S(t,{\bf p}_1) -S(t,{\bf p}_f)$, where the classical action is
$S(t,{\bf p}) = \frac{1}{2}\int dt[{\bf p} - {\bf A} _c(t)]^2$
with the potential of the corresponding classical laser field
being ${\bf A} _c(t)= \sqrt{U_p}[\hat{\epsilon} e^{-i\omega
t}+c.c.]$. Using the saddle-point approximation, we may find that
there are two moments $t_1$ and $2\pi/\omega- t_1$ that the
rescattering occurs within one optical cycle, which corresponds to
two classical trajectories; hence the interference fringes shown
in Fig.~\ref{f2} is due to the interference of these two classical
trajectories, where the minima are determined as $F=(2n+1)\pi/2$.
Although Eq.~(\ref{e9}) is so complex that there is no direct way
to show the relationship between the number of the interference
fringes and the system condition, the results in Fig.~\ref{f2}
indicate that some information can be derived from the
interference fringes. Figure \ref{f2} shows that the smaller the
$U_p/\omega$ and the smaller the incoming kinetic energy of the
rescattering electron, the fewer the interference fringes.

\begin{figure}[t]
\includegraphics[width=0.5\columnwidth]{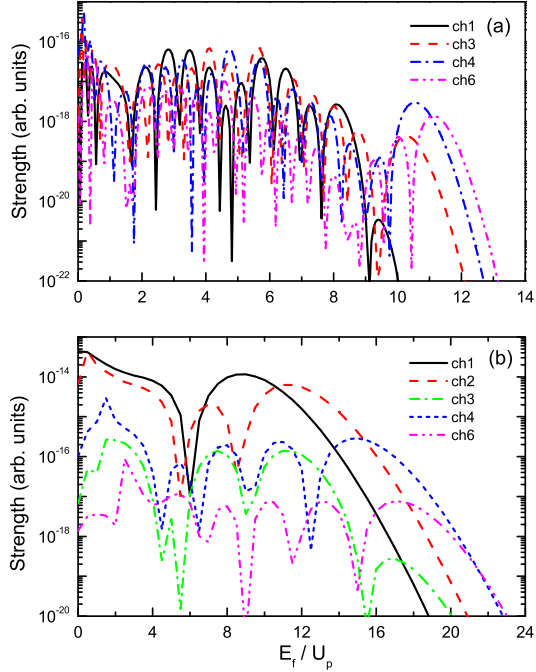} \vspace{-0.2cm}
\caption{(Color online) HATI spectra of H$_2^+$ for $\theta_f=0$
with $\omega =0.057$ (a) and $0.114$ (b).}\label{f3}
\end{figure}

Comparing the total HATI spectra in Fig.~\ref{f1} with the channel
HATI spectra in Fig.~\ref{f2}, one can see that, for the case of
low frequency, the total HATI spectrum is more complex than that
of channel 1; while for the case of high frequency, the total
spectrum agrees with that of channel 1. Since the total spectrum
is the coherent summation of the contributions from all ATI
channels, the above results indicate that the number of ATI
channels that contribute effectively to HATI decreases as $\omega$
increases. To confirm this finding, we present in Fig.~\ref{f3}
the HATI spectra with $\theta_f=0$ for different ATI channels for
the case of $\omega=0.057$ (a) and 0.114 (b) respectively.
Figure~\ref{f3} shows that the strength of HATI spectrum decreases
more rapidly with channel order for the case of higher frequency,
leading to a further simplification of the total HATI spectrum.

\begin{figure}[t]
\includegraphics[width= \columnwidth]{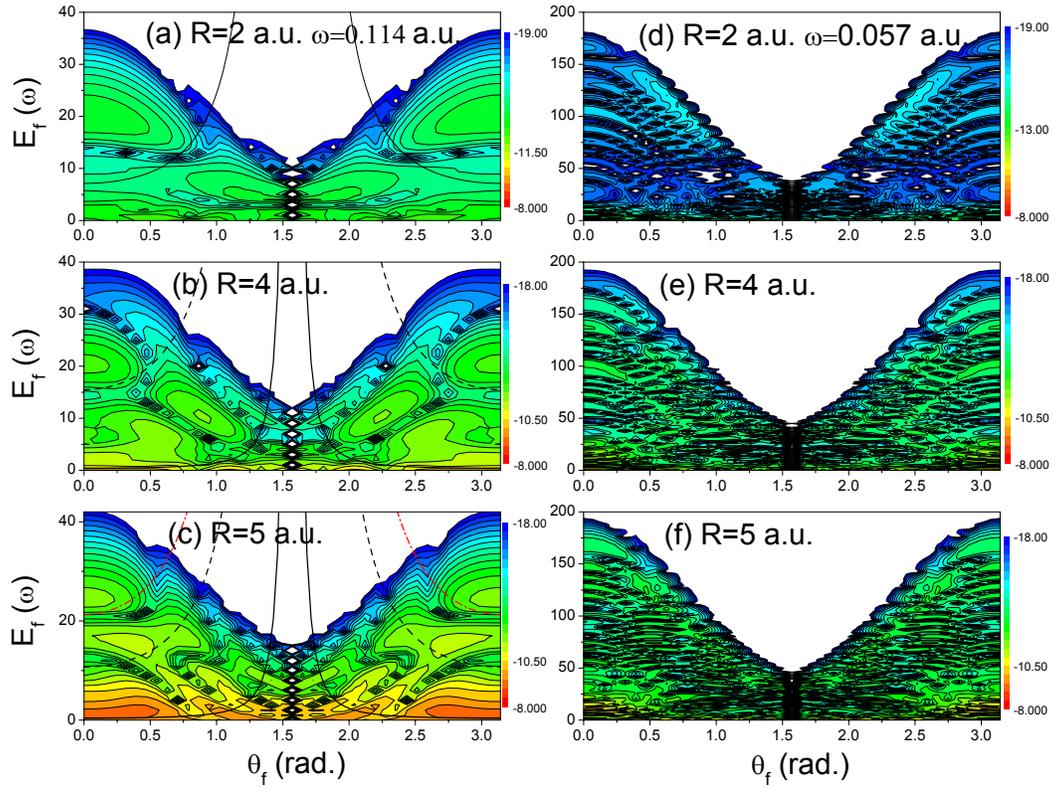} \vspace{-0.2cm}
\caption{(Color online) Angle-resolved HATI spectra of H$_2^+$
with the $\omega =0.114$ ((a)-(c)) and $0.057$ ((d)-(f)) for the
internuclear distance $R_0=2$ ((a) and (d)), 4 ((b) and (e)), and
5 ((c) and (f)). The results are plotted in log scale.}\label{f4}
\end{figure}

We now investigate the interference pattern of the HATI spectrum
for different geometry structures of the H$_2^+$ molecular ion. Using
the two-center potential of the molecule, the potential scattering
of the plane wave can be expressed as~\cite{8,9}£º
\begin{eqnarray}
\langle {\bf p}_f\mid U \mid {\bf p}_1\rangle \propto
\cos(\frac{R_0}{2}[p_f\cos \theta _f-p_1\cos \theta _1]),
\label{e14}
\end{eqnarray}
where $\theta_f$ is the angle between the molecular axis and the
final momentum ${\bf p}_f$ and $\theta_1$ is the angle between the
molecular axis and the momentum ${\bf p}_1$. Therefore, the
condition $p_f\cos \theta _f-p_1\cos \theta _1={(2n+1)\pi}/R_0$
corresponds to the destructive position in the HATI spectrum,
which contains the characteristic of the geometry structure of the
molecule. As an example, Fig.~\ref{f4} presents the HATI spectra
of H$_2^+$ with $R_0=2$ (a), 4 (b), and 5 (c) respectively and
$\omega=0.114$. For comparison, the corresponding spectra with
$\omega=0.057$ are also presented (Fig.~\ref{f4} (d-f)). One can
find that the HATI spectrum for the low frequency is hard to
clearly image the structure of the molecule because the background
interference is too complex. These results agree with the work by
Becker \emph{et al.}~\cite{9}, where the angle-resolved HATI
spectra shown in Figs.~6 and 7 of that paper are so complex that
the destructive curves are hard to be identified although these
curves are predicted (solid lines) by Eq.~(\ref{e14}). In
contrast, the molecular structure can be clearly imaged by the
HATI spectrum as the frequency increases, as shown in
Fig.~\ref{f4} (a-c). Furthermore, the pattern of the HATI spectrum
changes dramatically with $R_0$, and the destructive position can
be well predicted by the formula $p_f\cos \theta
_f+p_1=\frac{(2n+1)\pi}{R_0}$ ($n=1$ for solid black curve and
$n=2$ for dashed black curve) and $p_f\cos \theta
_f-p_1=\frac{(2n+1)\pi}{R_0}$ ($n=1$ for dot red curve).

In summary, we investigate the HATI of a molecule in a strong
laser field by the frequency-domain theory. The HATI can be
treated as a two-step transition: an ATI followed by a LAC. Based
on this viewpoint, It is found that the characteristic of a
molecule-laser system can be identified from the HATI spectrum:
the fingerprint of the molecular wave function lies in the ATI
transition, the geometry structure of the molecule lies in the LAC
transition with a phase factor which represents the interaction
between the ionized electron and the laser field. Furthermore, we
demonstrate that the HATI spectrum of high frequency laser can
clearly image the molecular structure of large internuclear
distance whereas it is hard to be identified for the low frequency
laser because of its complex nature of the interference pattern.

This research was supported by the National Natural Science
Foundation of China under Grant No. 60478031 and the 973 Research
Project under Grant No. 2006CB806003. ZCY was supported by NSERC
of Canada. BW thanks Biao Wu and Jing Chen for their helpful
suggestions.



\end{document}